\documentclass[aps,prl,twocolumn,groupedaddress,showpacs,reprint,superscriptaddress]{revtex4-1}

\usepackage{graphicx}
\begin{document}

%\preprint{}

\title{Deviation from Universality in Collisions of Ultracold $^6$Li$_2$ Molecules}


\author{Tout T. Wang}
\email[Email: ]{tout@physics.harvard.edu}
\affiliation{MIT-Harvard Center for Ultracold Atoms, Research Laboratory of Electronics, Department of Physics, Massachusetts Institute of Technology, Cambridge, MA 02139, USA}
\affiliation{Department of Physics, Harvard University, Cambridge, MA 02138, USA}

\author{Myoung-Sun Heo}
\altaffiliation[Current address: ]{Korea Research Institute of Standards and Science, Daejon 305-340, Korea}
\affiliation{MIT-Harvard Center for Ultracold Atoms, Research Laboratory of Electronics, Department of Physics, Massachusetts Institute of Technology, Cambridge, MA 02139, USA}

\author{Timur M. Rvachov}
\affiliation{MIT-Harvard Center for Ultracold Atoms, Research Laboratory of Electronics, Department of Physics, Massachusetts Institute of Technology, Cambridge, MA 02139, USA}

\author{Dylan A. Cotta}
\altaffiliation[Current address: ]{Department of Physics, University of Strathclyde, Glasgow G4 0NG, UK}
\affiliation{MIT-Harvard Center for Ultracold Atoms, Research Laboratory of Electronics, Department of Physics, Massachusetts Institute of Technology, Cambridge, MA 02139, USA}
\affiliation{D\'epartement de Physique, \'Ecole Normale Sup\'erieure de Cachan, 94235 Cachan, France}

\author{Wolfgang Ketterle}
\affiliation{MIT-Harvard Center for Ultracold Atoms, Research Laboratory of Electronics, Department of Physics, Massachusetts Institute of Technology, Cambridge, MA 02139, USA}


%\date{\today}

\begin{abstract}
Collisions of $^6$Li$_2$ molecules with free $^6$Li atoms reveal a striking deviation from universal predictions based on long-range van der Waals interactions. Li$_2$ closed-channel molecules are formed in the highest vibrational state near a narrow Feshbach resonance, and decay via two-body collisions with Li$_2$, Li, and Na. For Li$_2$+Li$_2$ and Li$_2$+Na, the decay rates agree with the universal predictions of the quantum Langevin model. In contrast, the rate for Li$_2$+Li is exceptionally small, with an upper bound ten times smaller than the universal prediction. This can be explained by the low density of available decay states in systems of light atoms [G. Qu\'em\'ener, J.-M. Launay, and P. Honvault, Phys. Rev. A \textbf{75}, 050701 (2007)], for which such collisions have not been studied before.
\end{abstract}

% http://www.aip.org/pacs/pacs2010/individuals/pacs2010_regular_edition/index.html
% 03.65.Nk 	Scattering theory
% 03.75.-b Matter waves
% 34.20.Gj Intermolecular and atom-molecule potentials and forces
% 34.50.-s 	Scattering of atoms and molecules
% 37.10.Pq 	Trapping of molecules
% 67.85.-d Ultracold gases, trapped gases
% 67.85.Lm 	Degenerate Fermi gases
%\keywords{}

\date{\today}

\maketitle

Recent advances in the preparation of ultracold samples of molecules are beginning to reveal how chemical reactions can occur in dramatically different ways at the quantum level compared to what happens in thermal ensembles \cite{QuemenerChemRev}. Beyond seminal experiments demonstrating the effects of quantum statistics and induced dipole moment on exothermic atom-exchange reactions of $^{40}$K$^{87}$Rb molecules \cite{OspQuantContReact}, there is a range of new possibilities to be explored \cite{QuemenerChemRev,KremsColdCont,BellUltracoldChem,ShapiroCohContr}, such as controlling collisions through magnetic/electric field tuning to access scattering resonances or level crossings, understanding whether reactions can depend sharply on the specific quantum state of the collision partners or on the details of short-range inter-particle interactions, characterizing the outgoing states of reaction products, demonstrating coherent control of reaction cross sections, and more.

There exists a simple, universal description for two-body inelastic collisions and chemical reactions of an ultracold molecule with another molecule or atom \cite{GaoUniv,JulUnivMolRates,QuemenerUnivers}. This quantum Langevin model assumes a large number of available exit channels in the short-range part of the interaction potential, leading to a unit probability of loss there, and leaving the decay rate dependent on only the long-range van der Waals interaction between collision partners. It has been validated in various experimental settings \cite{QuemenerChemRev}, involving heavier alkali molecules like Rb$_2$ \cite{WynarRb2}, Cs$_2$ \cite{StaanumCs2,ZahzamCs2}, KRb \cite{OspQuantContReact}, RbCs \cite{HudsonRbCs}, and LiCs \cite{DeiglLiCs}.

These universal collisions have a 100\% probability of loss at short-range and therefore do not depend on details of the interaction potential there, such as scattering resonances or reactivity determined by matrix elements between quantum states. From a chemistry standpoint, it is thus more interesting to search for examples of collisions that deviate from universality. Such deviations should be more prominent in systems with low mass and consequently a low density of available decay states \cite{GoulvenEmail,PaulEmail}, making $^6$Li$_2$, consisting of the lightest alkali atoms, a uniquely suitable experimental system. We observe that two-body collisions of Li$_2$ in the highest vibrational state with free Li atoms deviates sharply from universality, as reflected in an exceptionally small two-body decay coefficient. In contrast, the rates for both Li$_2$+Li$_2$ and Li$_2$+Na collisions are universal. A recent experiment inferred the rate of Li$_2$+Li decay from atomic three-body loss, but in a model-dependent way with uncertainty overlapping both our measurement and the universal prediction \cite{HazlettNarrowRes}.

To our knowledge this is the first experimental realization of collisions with ultracold molecules where loss is described by physics beyond universal long-range van der Waals interactions \footnote{We exclude weakly bound open-channel halo dimers near Feshbach resonances, to which the quantum Langevin model does not apply because their size is determined by the resonant two-body scattering length $a$, which can be much larger than the range of van der Waals interactions}. Earlier work by the Rice group reported a low decay rate for Li$_2$+Li$_2$ collisions \cite{StreckerLi2}, many orders of magnitude smaller than the universal prediction. In contrast, our measurement demonstrates that this rate is universal, addressing a puzzle that has been prominent for the last decade.

We sympathetically cool fermionic $^6$Li with bosonic $^{23}$Na in a magnetic trap, as described in our earlier work \cite{HeoNaLi}. The number balance as well as the final temperature of the atoms can be adjusted by changing the Na evaporation endpoint and the initial loading times from the two atomic beams. We can completely evaporate Na, leaving a pure Li gas, or interrupt evaporation part-way to obtain a Na+Li mixture. At the end of magnetic trap evaporation, the atoms are transferred into a single-beam optical dipole trap with $5$~W power and wavelength $1064$ nm. Then we spin-flip Li and any remaining Na atoms ($\left| F, m_F \right> =  \left| 3/2, 3/2 \right>\rightarrow \left| 1/2, 1/2 \right>$ and $\left| 2, 2 \right>\rightarrow\left| 1, 1 \right>$ respectively) with simultaneous Landau-Zener radio-frequency (rf) sweeps at $15$ G. An equal superposition of the two lowest Li hyperfine states $\left| 1 \right>$ and $\left| 2 \right>$ (corresponding to $\left| 1/2, 1/2 \right>$ and $\left| 1/2, -1/2 \right>$ at low field) is prepared by two identical, non-adiabatic rf sweeps at 300 G, separated by 10 ms. The second sweep compensates for small imbalances resulting from the first. After holding for a further 500 ms, the superposition becomes an incoherent mixture of $\left| 1 \right>$ and $\left| 2 \right>$ states \cite{GuptaRF, Varenna2008}. We then further evaporatively cool the Li (with or without Na present), to $T/T_F=0.2$, and transfer the atoms into a second, more weakly confining single-beam optical dipole trap parallel to the first one, with trap frequencies $(\nu_z,\nu_r)$ = $(21,480)$ Hz. Optimized formation of Li$_2$ takes place in this second trap.

\begin{figure}\centering
\includegraphics[scale=0.55]{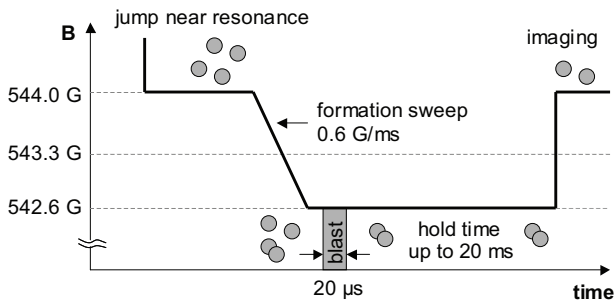}
\caption{Experimental sequence for molecule formation. After jumping near the $B=543$ G resonance from above, we sweep the magnetic field $B$ at 0.6 G/ms across a region 1.4 G wide. Immediately after the sweep, a short pulse of resonant imaging light removes free Li from the trap, leaving a pure gas of Li$_2$. After a variable hold time, remaining molecules are detected by jumping back above resonance and imaging dissociated free atoms, or $B$ can be kept below resonance to confirm that Li$_2$ is invisible to the imaging light.}\label{fig:sweeps}
\end{figure}

Initial molecule formation experiments are done with Li $\left| 1 \right>$ and $\left| 2 \right>$ states only, without Na present. The initial number in each state is $2\times 10^6$, corresponding to a peak density of $4\times 10^{12}$ cm$^{-3}$.  A magnetic field sweep \cite{KohlerProdMol} across the narrow $^6$Li Feshbach resonance at $B=543$ G \cite{OHaraZeroCross,StreckerLi2,HazlettNarrowRes,ChinFesh} (Fig.~\ref{fig:sweeps}) converts the atoms into diatomic molecules.  These are closed-channel molecules in the highest vibrational state of the $^1\Sigma^+_g$ potential.

\begin{figure}\centering
\includegraphics[scale=0.42]{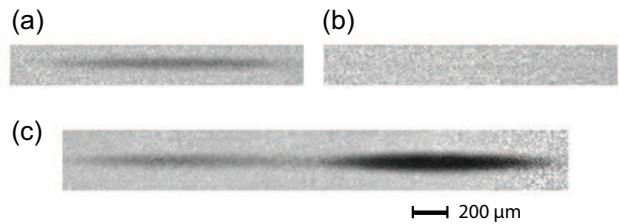}
\caption{Images of Li$_2$ molecules. Absorption images of (a) molecules dissociated into free atoms above the 543 G Feshbach resonance, (b) molecules held below resonance where they are invisible to the imaging light, and (c) separation of molecules (left) and atoms (right) in a magnetic field gradient.}\label{fig:gradient}
\end{figure}

As in our previous work \cite{HeoNaLi}, we unambiguously observe the signature of molecule formation by applying a short blast of resonant imaging light to remove free Li in the $\left| 1 \right>$ state from the trap. The light mass of Li means that it will be ejected from the trap after a single recoil, and moreover at 543 G the imaging transition is cycling, so a pulse duration of 20 $\mu$s is sufficient to leave no trace of $\left| 1 \right>$ atoms.  After the blast, imaging the $\left| 1 \right>$ state while keeping $B$ below resonance gives a negligible signal, since there Li$_2$ is invisible to the imaging light. After switching $B$ above resonance to dissociate Li$_2$, we image atoms in the $\left| 1 \right>$ state as a measure of the molecule number, which for our optimized sweep parameters gives a formation fraction of $10\%$ or a molecule number of $2\times 10^5$. A second, independent confirmation of molecule formation is obtained by turning on a magnetic field gradient of 10 G/cm for 6 ms while holding in-trap below resonance, which pushes free Li atoms away while leaving the spin-singlet molecules unaffected (Fig.~\ref{fig:gradient}).

\begin{figure}\centering
\includegraphics[scale=0.65]{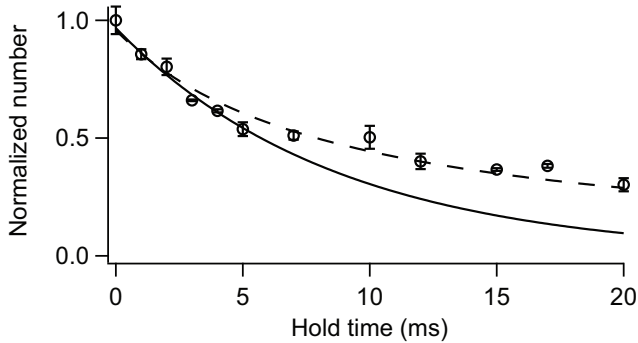}
\caption{Lifetime of a pure sample of Li$_2$ molecules without free atoms present. The solid line is an exponential fit up to 5 ms hold time, giving a decay time of 8.7(5) ms, while the dashed line is a fit to a full two-body decay function.}\label{fig:shallowlifetime}
\end{figure}

For molecule decay measurements, we use two consecutive blasts of imaging light, resonant with states $\left| 1 \right>$  and $\left| 2 \right>$ respectively, to remove free atoms in both hyperfine states from the trap immediately after the molecule formation sweep. This leaves a pure sample of Li$_2$ molecules, which undergoes rapid initial decay from their vibrationally excited state, slowing down with increasing hold time in a way that is consistent with two-body decay from molecule-molecule collisions (Fig.~\ref{fig:shallowlifetime}). The non-exponential nature of the decay rules out that lifetimes are limited by off-resonant excitations from the trapping laser, and we have also checked, by holding Li$_2$ in trap at up to 8 G below resonance, that decay rates outside the coupling region around resonance are independent of magnetic field, as expected. We determine a two-body decay coefficient $\beta_\mathrm{Li_2+Li_2}=6(2)\times 10^{-10}$ cm$^3$/s, with the uncertainty dominated by systematic errors in determining densities. The quoted result comes from averaging multiple data sets, including those from a different crossed-beam trap geometry with higher initial densities.

\begin{figure}\centering
\includegraphics[scale=0.65]{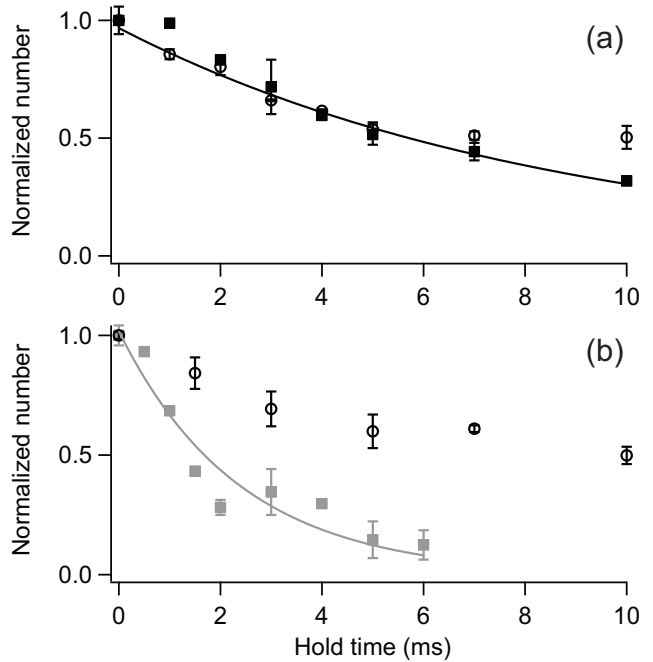}
\caption{Lifetime of Li$_2$ in the presence of free atoms. (a) Decay with Li (solid squares) is almost identical to the decay of the pure Li$_2$ gas (open circles). (b) Decay with a similar initial density of Na (solid squares) is significantly faster than the decay of the pure Li$_2$ gas (open circles). }\label{fig:atoms}
\end{figure}

Applying the $\left| 1 \right>$  and $\left| 2 \right>$ state blast beams at the end of the hold time instead of immediately after the molecule formation sweep allows us to measure molecule decay in the presence of free Li atoms. The Li density is much higher than that of the molecules, so the presence of the atoms should significantly increase the decay rate. Surprisingly, we find that Li gives only a small, non-observable contribution to the decay [Fig. \ref{fig:atoms}(a)] when compared to the decay from Fig. \ref{fig:shallowlifetime}, corresponding to an upper bound $\beta_\mathrm{Li_2+Li}<5\times 10^{-11}$ cm$^3$/s. There should be no Pauli suppression of collisions for closed-channel Li$_2$ molecules with Li \cite{PetrovFermiDimer,RegalLifetime}. We confirm this by checking that there is no enhancement of the decay after spin-flipping one component of Li from $\left| 2 \right>$ to $\left| 3 \right>$ ($\left| F, m_F \right> =  \left| 3/2, -3/2 \right>$).

When Na instead of Li atoms are trapped with Li$_2$, significant enhancement of the loss rate is observed for similar initial atomic densities [Fig.~\ref{fig:atoms}(b)], corresponding to $\beta_\mathrm{Li_2+Na}=4(1)\times 10^{-10}$ cm$^3$/s. Na and Li only interact weakly \cite{OberNaLi}, thus the presence of Na has a negligible effect on Li$_2$ molecule formation. The mixture is produced with the same temperature and initial Li$_2$ density as in decay measurements done without Na.

Two-body decay of molecules is described by
\begin{equation}\label{eq:twobody}
\frac{\dot n_\mathrm{Li_2}}{n_\mathrm{Li_2}}=-\beta_\mathrm{Li_2+Li_2}n_\mathrm{Li_2}-\beta_\mathrm{Li_2+Li}n_\mathrm{Li}-\beta_\mathrm{Li_2+Na}n_\mathrm{Na}
\end{equation}
where $n$ represents local densities. Experimentally, we measure the decay of total molecule number N$_\mathrm{Li_2}$ rather than n$_\mathrm{Li_2}$, so Eq.~(\ref{eq:twobody}) can be written, assuming separate Gaussian density distributions for Li$_2$, Li and Na
\begin{equation}\label{eq:twobody2}
2^{3/2}\frac{\dot N_\mathrm{Li_2}}{N_\mathrm{Li_2}}=-\beta_\mathrm{Li_2+Li_2}\tilde{n}_\mathrm{Li_2}-\beta_\mathrm{Li_2+Li}\tilde{n}_\mathrm{Li}-\beta_\mathrm{Li_2+Na}\tilde{n}_\mathrm{Na}
\end{equation}
with $\tilde{n}$ denoting peak in-trap densities. The various two-body decay coefficients $\beta$ can thus be extracted by fitting exponential decay rates at short hold times and normalizing by the initial peak densities. The factor $2^{3/2}$ accounts for the variation of density across the trap. The effect of deviations of density profiles from Gaussian is much smaller than the quoted uncertainties for $\beta$.

Full expressions for trapped ideal Bose(Fermi) gases in the local density approximation are used to calculate peak densities $\tilde{n}_\mathrm{Na(Li)}=\pm(\frac{m k_B T}{2\pi\hbar^2})^{3/2}\mathrm{Li}_{3/2}(\pm z)$, where $m$ is the mass of the Na(Li) atom, $T$ is the temperature of the gas, and $\mathrm{Li}_n(z)$ is the $n$-th order Polylogarithm \cite{Varenna2008}.  We determine $T$ before the molecule formation sweep by fitting Li time-of-flight expanded 2D column density profiles with the fugacity $z=e^{\beta\mu}$ as a free parameter, giving $T/T_F=0.2$ or $T=400$ nK. After the molecule formation sweep, $\tilde{n}_\mathrm{Li}$ is lower by a factor of two compared to before the sweep, despite a molecule conversion efficiency of only $10\%$, because many more atoms are associated into molecules that are lost via collisions with other molecules in the time it takes to complete the sweep.

For the Li$_2$ density, the simplest assumption is that the density distribution is proportional to the Li density profile before the sweep, meaning that $\tilde{n}_\mathrm{Li_2}$ can be estimated from the ratio of total numbers
\begin{equation}\label{eq:ratio}
\tilde{n}_\mathrm{Li_2}=\tilde{n}_\mathrm{Li} N_\mathrm{Li_2}/N_\mathrm{Li}
\end{equation}
and likewise for the reduced $\tilde{n}_\mathrm{Li}$ after the sweep.  This is valid if we neglect the complicated density-dependence of molecule formation efficiency \cite{KohlerProdMol} and assume that Li atoms are selected at random from the Fermi sea to form molecules that do not have time to reach thermal equilibrium. We improve on this by accounting for the effect of equilibration. Molecules form with the same average center-of-mass kinetic energy as the free atoms, but twice the potential energy (due to their larger polarizability). Assuming that equilibration distributes this excess energy among all the degrees of freedom (according to the Virial theorem applied to harmonic traps), the cloud width is rescaled by $\sqrt{3/4}$. In our experiment, Li$_2$ does not equilibrate along the weak axial trapping direction, so the radial cloud diameter is instead rescaled by $\sqrt{7/10}$. This implies that Eq.~(\ref{eq:ratio}) underestimates the peak Li$_2$ density by about 30\%, which is within the quoted uncertainty of our results, but we nevertheless include the correction in the analysis. This correction does not apply to $\tilde{n}_\mathrm{Li}$ after the sweep \footnote{Random loss from a Fermi-Dirac distribution leaves the root-mean-square size of equilibrated cloud unchanged}.

\begin{figure}\centering
\includegraphics[scale=0.8]{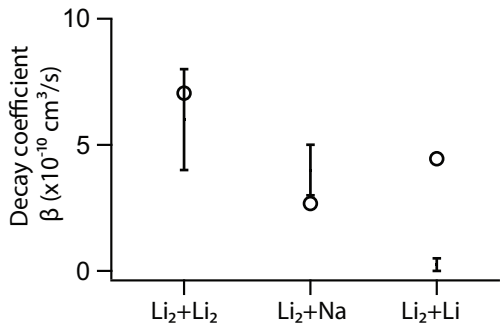}
\caption{Predictions for two-body inelastic decay coefficients from the universal model (open circles) compared to experimental measurements (vertical bars).}\label{fig:univ}
\end{figure}

Figure~\ref{fig:univ} shows how $\beta_\mathrm{Li_2+Li}$ deviates sharply from from the universal prediction while $\beta_\mathrm{Li_2+Li_2}$ and $\beta_\mathrm{Li_2+Na}$ are universal. In the quantum Langevin model for universal collisions of ultracold molecules, the assumption of total loss at short-range leaves the decay rate dependent only on the long-range van der Waals interaction via \cite{JulUnivMolRates,QuemenerUnivers}
\begin{equation}\label{eq:univ}
\beta=g \frac{4\pi\hbar\overline{a}}{\mu},\quad \overline{a}\approx 0.48(\frac{2\mu C_6}{\hbar})^{1/4}
\end{equation}
where the prefactor $g=1(2)$ for (in)distinguishable collision partners accounts for wave-function symmetrization, $\mu$ is the reduced mass of the two-body system, and with a length scale $\overline{a}$ for van der Waals interactions expressed in terms of the $C_6$ coefficient. The weak $C_6^{1/4}$ dependence means that variations of $\beta$ come mainly from differences in $\mu$. For collisions involving weakly bound molecules, $C_6$ can be approximated as the sum of the corresponding coefficients for all combinations of atom pairs involved \cite{DeiglLiCs}, with the atomic $C_6$ coefficients taken from calculations \cite{DereVdW,MitroyVdW}.

The observed deviation from universality in Li$_2$+Li collisions can be explained using a full close-coupling quantum calculation that predicts a two-body coefficient of $5\times 10^{-11}$ cm$^3$/s for Li$_2$ in the least bound triplet $^3\Sigma^+_u$  state, and can be interpreted as a depletion mechanism \cite{QuemenerDepletion}. A similarly low value can be inferred for the least bound singlet $^1\Sigma^+_g$ state in our experiment from the same mechanism \cite{GoulvenEmail}. This effect is absent for Li$_2$+Na, Li$_2$+Li$_2$, and collisions involving heavier alkali molecules, because these all have a higher density of decay states in the exit channel \cite{QuemenerChemRev,QuemenerDepletion,GoulvenEmail}. The effect is also only present for the highest vibrational state, in which the atoms spend the longest time near the outer turning point of the van der Waals potential, where they do not have sufficient kinetic energy to transfer to a collision partner in a vibrational relaxation process. Li$_2$ in the second highest vibrational state is expected to have an order of magnitude larger $\beta_\mathrm{Li_2+Li}$ \cite{QuemenerDepletion}. Confirming this prediction requires using a two-photon Raman transition to change the vibrational state.

Closed channel molecules formed around a Feshbach resonance in the highest vibrational state have size on the order of the van der Waals length $\overline{a}$ \cite{QuemenerChemRev,ChinFesh}, while the universal model assumes a clean separation between loss processes at short range and the long-range van der Waals interaction. Our measurements show that Li$_2$+Li$_2$ and Li$_2$+Na collisions are still well described by universal predictions. This follows previous work, both theoretical \cite{GaoUniv} and experimental \cite{StaanumCs2,HudsonRbCs}, indicating that decay rates are mostly independent of the vibrational quantum number of the molecule. Decay of other molecules like Na$_2$ \cite{MukaiNaDissoc,YurovskyNa2} and Cs$_2$ \cite{ChinCs2} from their highest vibrational states also show fair agreement with universal predictions.

Our work addresses the puzzle of long lifetimes of Li$_2$ molecules observed by the Rice group \cite{StreckerLi2}. Long lifetimes of molecules consisting of a pair of fermions were crucial for the exploration of the BEC-BCS crossover \cite{Varenna2008}. While the observation of long lifetimes near a broad Feshbach resonance \cite{SalomonLi2,JochimMolec} was quickly explained in terms of Pauli suppression of collisions involving open-channel-dominated fermion pairs \cite{PetrovFermiDimer}, the other observation near a narrow resonance has remained unexplained, since Pauli suppression should be absent for closed-channel molecules outside the narrow coupling region around resonance \cite{RegalLifetime}. Our measurements demonstrate that the lifetimes of these molecules are short (Fig.~\ref{fig:shallowlifetime}) due to Li$_2$+Li$_2$ collisions being universal.

We are not aware of any differences between the two experiments that can explain the discrepancy. The atomic densities and temperatures in both experiments are comparable. In addition to the results presented above, where molecule lifetimes were measured at a constant magnetic field, we have reproduced the magnetic field sweep during the hold time used at Rice \cite{StreckerThesis} and found no significant lifetime enhancement. A reviewer suggested that residual coherence between atoms in $\left| 1 \right>$ and $\left| 2 \right>$ states in the Rice experiment could lead to longer lifetimes. Coherence of the atoms can affect molecule-atom collisions, but not molecule-molecule collisions for which the rate coefficients deduced from the two experiments differ by more than a factor of a hundred \footnote{The reported densities and lifetimes correspond to $\beta<2\times 10^{-12}$ cm$^3$/s for both Li$_2$+Li$_2$ and Li$_2$+Li collisions in the Rice experiment}. The Rice results are also incompatible with recent work inferring $\beta_\mathrm{Li_2+Li}$ from three-body atomic loss \cite{HazlettNarrowRes}, which has not been pointed out before.  Note that the Rice experiment deduced the presence of molecules from differences in Li atom numbers \cite{StreckerLi2}, whereas our experiment identifies Li$_2$ molecules in addition by two highly specific methods (magnetic field gradient separation, and  survival of molecules when Li atoms are removed by resonant light, both shown in Fig.~\ref{fig:gradient}).

In summary, we have observed that, for Li$_2$ molecules in the highest vibrational state formed around a narrow Feshbach resonance, Li$_2$+Li collisions deviate sharply from the universal predictions of the simple quantum Langevin model, while Li$_2$+Li$_2$ and Li$_2$+Na collisions are universal. This is the first example of collisions involving ultracold molecules with loss determined by physics beyond long-range van der Waals interactions.

%%%%%%%%%%%%%%%%%%%%%%%%%%%%%%%%%%%%%%%%%%%%%%%%%%%%%%%%
%% ACKNOWLEDGEMENTS
\begin{acknowledgments}
This work was supported by an AFOSR MURI on Ultracold Molecules, an ARO MURI on Quantum Control of Chemical Reactions, by the NSF and the ONR, and by ARO Grant No. W911NF-07-1-0493, with funds from the DARPA Optical Lattice Emulator program. T.T.W. acknowledges additional support from NSERC. We are grateful to G. Qu\'em\'ener and P.S. Julienne for many valuable discussions.
\end{acknowledgments}

%\bibliography{Li2}


\end{document}